# FOCCoS for Subaru PFS

Antonio Cesar de Oliveira[1], Ligia Souza de Oliveira[1], Marcio Vital de Arruda[1], Jesulino Bispo dos Santos[1], Lucas Souza Marrara[1], Vanessa Bawden de Paula Macanhan[1], João Batista de Carvalho Oliveira[1], Rodrigo de Paiva Vilaça[1], Tania Pereira Dominici[1], Laerte Sodré Junior[2], Claudia Mendes de Oliveira[2], Hiroshi Karoji[3], Hajime Sugai[3], Atsushi Shimono[3], Naoyuki Tamura[4], Naruhisa Takato[4], Akitoshi Ueda[4],

1- MCT/LNA –Laboratório Nacional de Astrofísica, Itajubá - MG - Brazil
2- IAG/USP – Instituto Astronômico e Geografico/ Universidade de Sao Paulo - SP – Brazil
3- IPMU University of Tokyo, Institute for the Physics and Mathematics of the Universe
4- Subaru Telescope/ National Astronomical Observatory of Japan

## ABSTRACT

The Fiber Optical Cable and Connector System (FOCCoS), provides optical connection between 2400 positioners and a set of spectrographs by an optical fibers cable as part of Subaru PFS instrument. Each positioner retains one fiber entrance attached at a microlens, which is responsible for the F-ratio transformation into a larger one so that difficulties of spectrograph design are eased. The optical fibers cable will be segmented in 3 parts at long of the way, cable A, cable B and cable C, connected by a set of multi-fibers connectors. Cable B will be permanently attached at the Subaru telescope. The first set of multi-fibers connectors will connect the cable A to the cable C from the spectrograph system at the Nasmith platform. The cable A, is an extension of a pseudo-slit device obtained with the linear disposition of the extremities of the optical fibers and fixed by epoxy at a base of composite substrate. The second set of multi-fibers connectors will connect the other extremity of cable A to the cable B, which is part of the positioner's device structure. The optical fiber under study for this project is the Polymicro FBP120170190, which has shown very encouraging results. The kind of test involves FRD measurements caused by stress induced by rotation and twist of the fiber extremity, similar conditions to those produced by positioners of the PFS instrument. The multi-fibers connector under study is produced by *USCONEC* Company and may connect 32 optical fibers. The tests involve throughput of light and stability after many connections and disconnections. This paper will review the general design of the FOCCoS subsystem, methods used to fabricate the devices involved and the tests results necessary to evaluate the total efficiency of the set.

## 1. INTRODUCTION

FOCCOS subsystem, Figure 1, is defined by 4 interfaces:

1. Telescope (External Physical Interface):
• Cable route & Telescope
The route will be defined taking in account the telescope structure, the available places for the parts of the instrument and the movement of the telescope. (TBD)

2. Spectrograph (External Optical Interface):
• Pseudo slit & Spectrograph
For 4 spectrographs we need 4 pseudo slits. Each one should be 140mm long and holds 600 fibers, so the center-to-center fiber spacing is 230 microns. The fiber OD is 190 microns.

3. Cobra (External Optical Interface):
• Fiber arms & HSC
Each COBRA positioner unit will contain a fiber arm that holds one fiber whose positions, are articulated to be positioned over a small region of sky. This fiber needs to be polished and secured to the COBRA unit in a robust, but stress-free manner, which retains the intrinsic FRD of the fibers themselves. The fiber extremity requires a microlens glued to accept the fast (~f/2.3) beam input from the PF corrector with minimal throughput loss.

4. Connectors (Internal Optical Interface):
• Cable A & Cable B & Cable C
The modules of connectors are required to perform with optimal efficiency through a large number (TBD) of connect/disconnect cycles and are to be made so that there is no impairment to the surface quality of the fibers themselves. It will be necessary to have connection with all cables, cable A to cable B and cable B to cable C. The connector under study is from USCONEC, which was successfully used in the Apogee spectrograph for the SDSS.

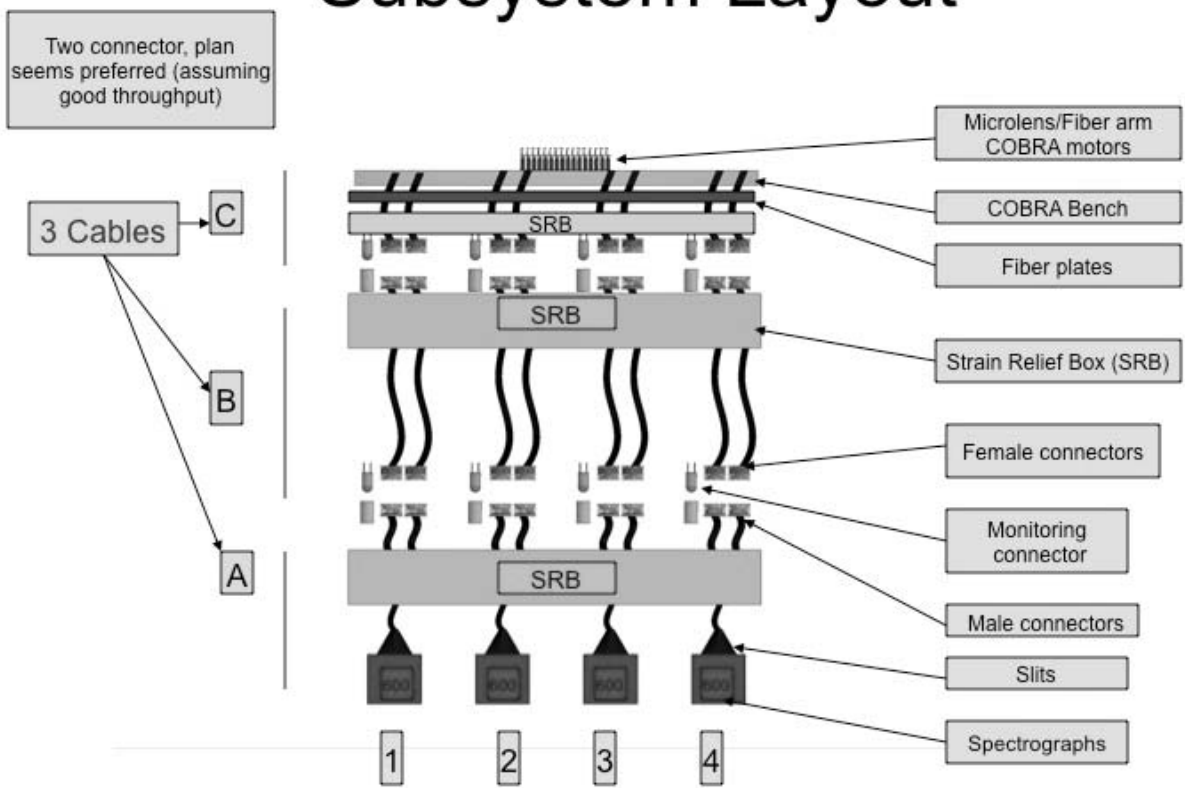

**Figure 1:** FOCCoS Layout subsystem showing the interfaces

All optical interfaces (OI), are involving a bundle of optical fibers divided in manageable groups. Physical interfaces (PI) refer to all other non-optical interfaces. Each group is encased by protective tubes and clamped along their route. The route starts at the COBRA positioner through a tower cell connectors (80 small unit Apogee) and then across the platform veins and down the telescope tube and through again the 8 Gang connectors to finish at the spectrograph.

## 2. CABLE SYSTEM

The cable system is sequentially divided in cable A, cable B and cable C, which are connected together by multifibers connectors. To protect the optical fibers from excessive strain or damage during operation and installation, a system of conduit tubes, furcation tubes, worm tubes and polyamide tubes will be used. The conduit tube is outer tube, which gives full protection to a group of furcation, or segmented tubes containing the optical fibers. In general they are constructed by steel rings covered by plastics or steel mesh as shown in the Figure 2. The furcation tube, Figure 3, is a set of double tubes constituting an external tube of plastic and an internal tube of PTFE within which several fibers are contained. Between the internal and external tubes there are Kevlar fibers. Furcations tubes are generally used to protect groups of optical fibers. The Kevlar and the external tube, of the Furcation tube are used to clamp a metallic termination. Segmented tubes, or worm tubes, Figure 4, are constructed of only one tube of PTFE plastic. However, the internal surface is extremely smooth, specially designed to accommodate optical fibers. The external surface is segmented, as shown. This kind of tube is good for allow the fibers to spin freely within the tube at the termination points. Polyamide tubes, Figure 5, are used as strain relief tube to prevent mechanical stress, and hence FRD, occurring at the point where the fiber enters the ferrule. The conduit contains the furcation tubes, each of which contain and protect approximately 30 optical fibers. The termination of each furcation tubes has a metallic termination point, which holds the furcation tube onto a plate. This plate is attached in the extremity of the protective conduit which itself is attached on the structure of the telescope.

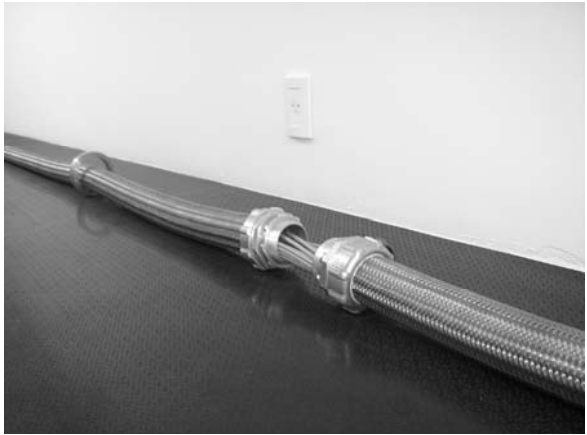

**Figure 2-** External conduit tube, segmented to receive the compensation box.

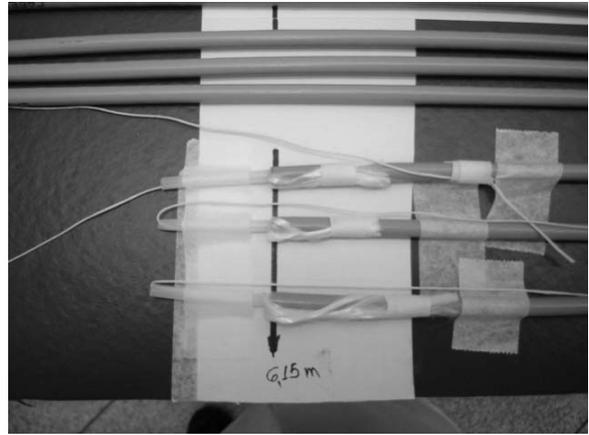

**Figure 3-** Arrangement of furcation tubes

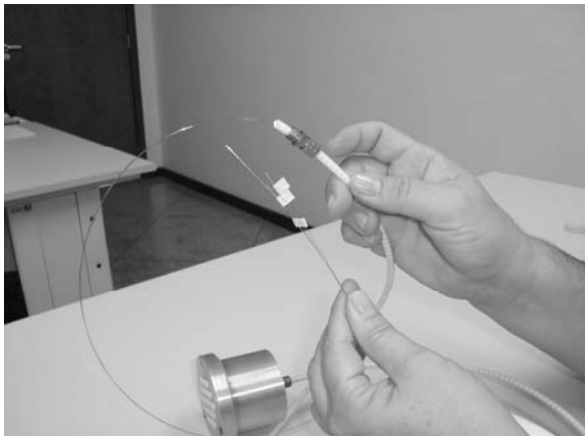

**Figure 4-** Worm (segmented) tubes

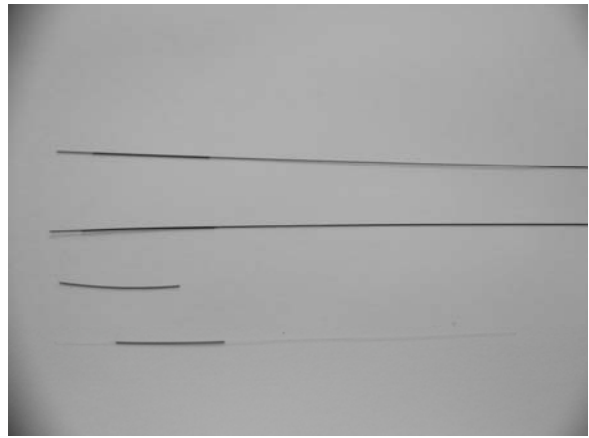

**Figure 5-** Polyamide tube

## 2.1 Cable A

Cable A, will be the cable installed at the spectrograph side, including the slit system. The composition of this cable is basically small conduit tubes, containing furcation tubes with groups of optical fibers to be distributed for 4 pseudo slits devices and to feed the spectrographs sets. Cable A is a set of parallels cables starting in the slit device, crossing a distribution box and finishing at the connectors system, Figure 6.

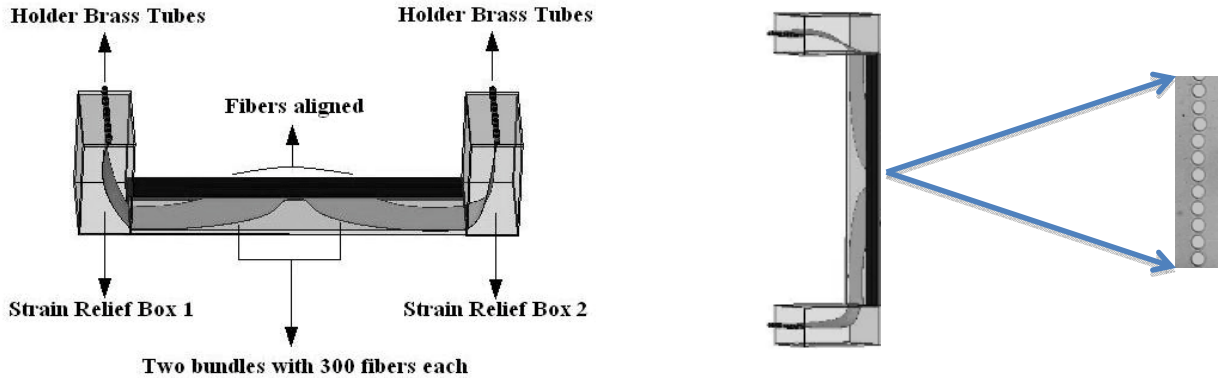

**Figure 6:** Fiber Device termination is one of the extremities of the cable A. Segmented tubes, starting in the holder brass, conduct the fibers at the other extremity of the cable A, called optical bench connectors.

### 2.1.1 Slit Device

The slit device is a monoblock, made and machined in composite, Figure 7, with all fibers disposed in a circular arrangement. Each one is 140 mm long and holds 600 fibers, so the center-to-center fiber spacing is 230 microns. The fiber OD is 190 microns. In this case, the best option is to use a mask of precision. The device assembled with the correct gap need to be made with two twin masks as a shown in Figure 8. This is a metal mask very thin obtained by a technique called electro formation. The mask obtained by this way can be configured to have holes with specifics diameters and pits, with errors around 1 micron in the diameter and in the position of the holes. This technique can produce a metal nickel plate with 200 microns of thickness and the procedure is very cheap. It is possible to obtain micro holes with the diameter exactly one or two microns larger than the diameter of the used fiber. Four slit devices will be constructed, each one containing 600 fibers aligned through holes of 190 microns in diameter with a pitch of 230 microns. The device is immersed in a container, with EPOTEK 301-2, constructed using plates of PTFE doped with graphite. After dry it is very easy to remove the frontal plate without offer any damage to the block or the fibers. The polishing process consists of: 1) to cut the end of the block and removal of the excess of glue with 2000 and 2500 grit emery paper; 2) initial lapping with 6 micron diamond slurry on a copper plate and 3) a second lapping with 1 micron diamond slurry on a tin-lead plate is used until the complete removal of the frontal mask of precision. The final polishing is made with colloidal silica solution on a chemical cloth with 0.01 micron. The Figure 9 shows the slit system during the assembly.

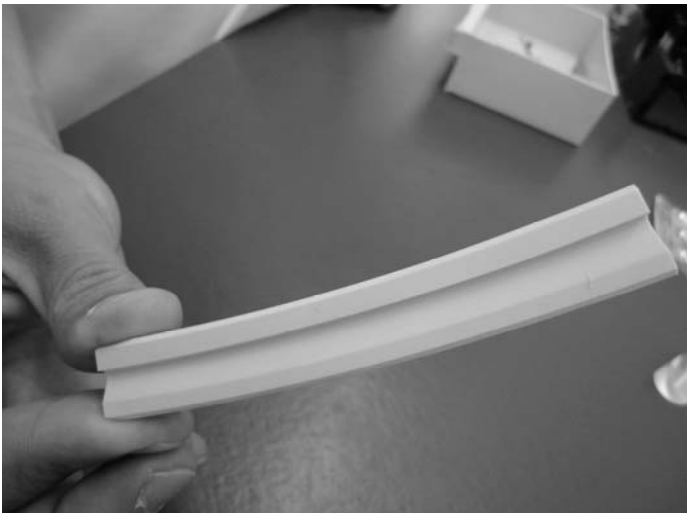

**Figure 7:** Monoblock made and machined to be the slit block

A meniscus lens whose inner radius R1 is the same as the radius of the fiber slit and whose outer radius R2, longer to make the lens negative, with AR coating in both sides may be glued against the slit block to protect the fibers extremities and avoid scratches or damages.

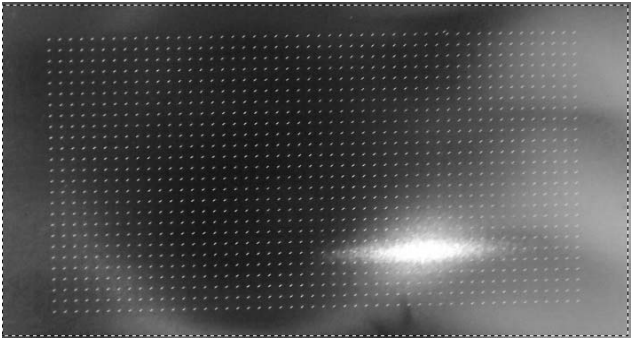

**Figure 8:** Photo of the mask of precision used to obtain the slit with correct distribution and gap between the optical fibers.

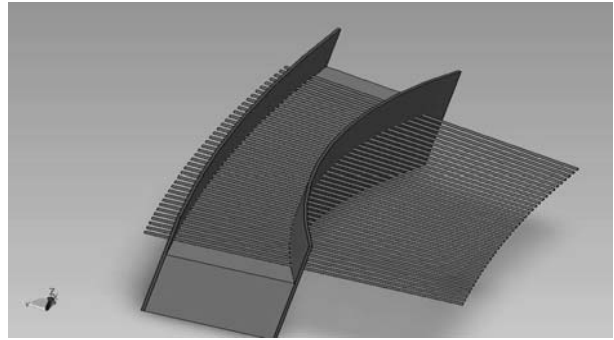

**Figure 9:** Schematic of the curved slit during the assembly. The fibers do not touch the base made with composite.

## 2.2 Cable B

Cable B will be the cable permanently installed at the telescope structure, composing the longest cable of the system, with around 40 meters. This cable is basically a conduit tube containing several plastic tubes inside (furcation tubes) through which the groups of optical fibers are protected, Figure 10. The routing of the cable is either looped across to the "Great Wall", within which the spectrographs are housed, or through the cable warp of the elevation axis. The cable will interface directly with the telescope and dome structure from the TES vanes to the Great Wall and it needs to define minimal stress on the fibers during telescope pointing and field rotation of the TES.

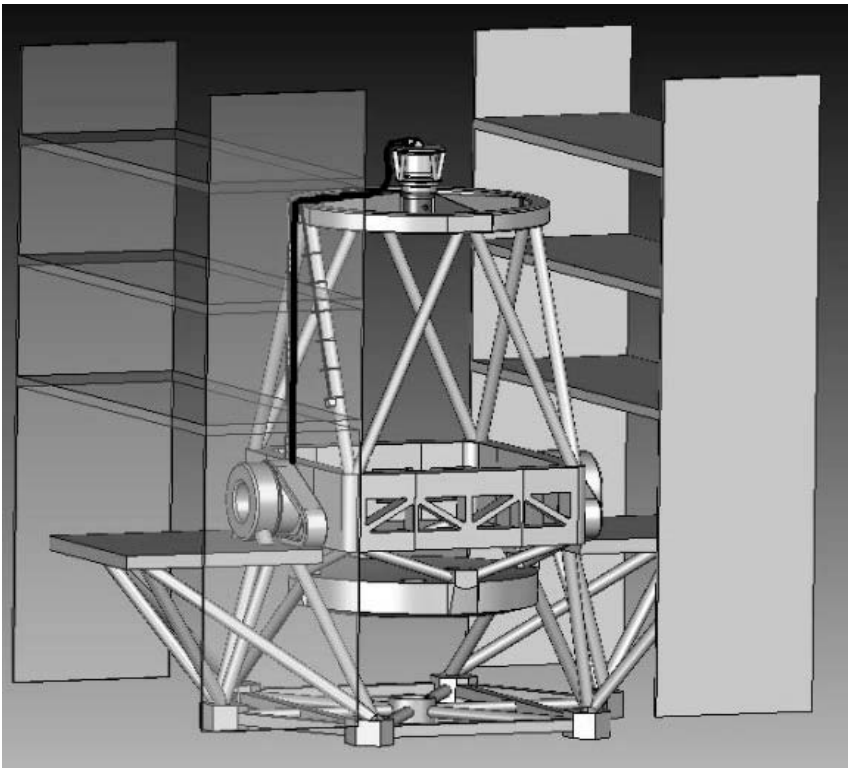

**Figure 10:** Possible disposition of the Cable B clamped at the telescope structure

Changes in the length between local fiber securing points will be accommodated through the use of constant force springs. A set of strain relief boxes, Figure 11, will be located at the extremities of the cable B, close to the spectrograph system and close to the top end system. This defines enough free length of fibers out of the protection tubes to facilitate the manipulation of the fibers during the construction and the polishing procedures of the cable. The cable B, at the top end side, will be separated in two parts to conduct fibers only by the furcation's tubes, willing in combs on both laterals of the spider plate, Figure12.

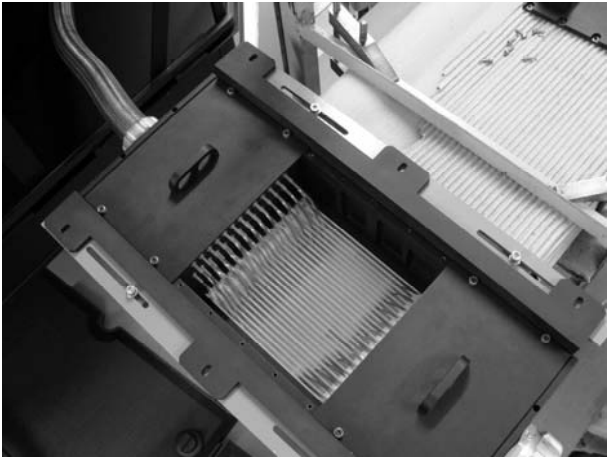

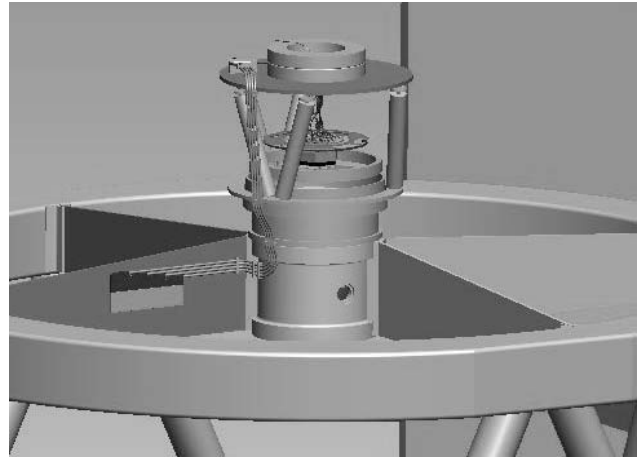

**Figure 11:** Strain relief box in study to be used in both extremities of the Cable B.

**Figure 12:** Combs of tubes containing optical fibers, in one of sides of the spider plate on the top end of the telescope.

## 2.3 Cable C

Cable C, will be the cable installed at the top end devices, the COBRA motors systems. It will be composed by a set of segmented tubes articulated inside of the PFI chamber, Figure 13. Cable C is a set of parallels cables that start in the fiber arm devices, crossing the COBRA plate, fiber plates, strain relief boxes and finishing at the connectors system, Figure 14.

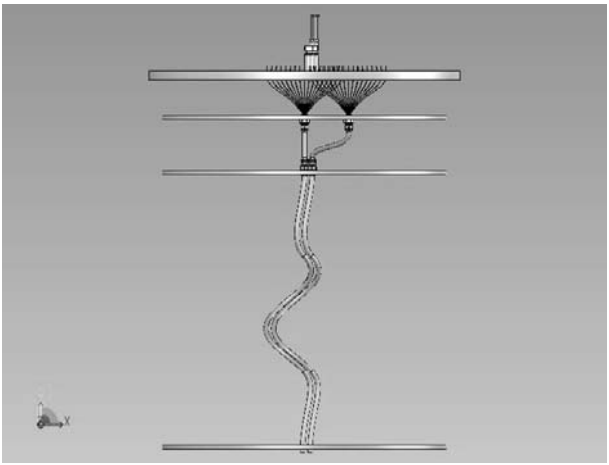

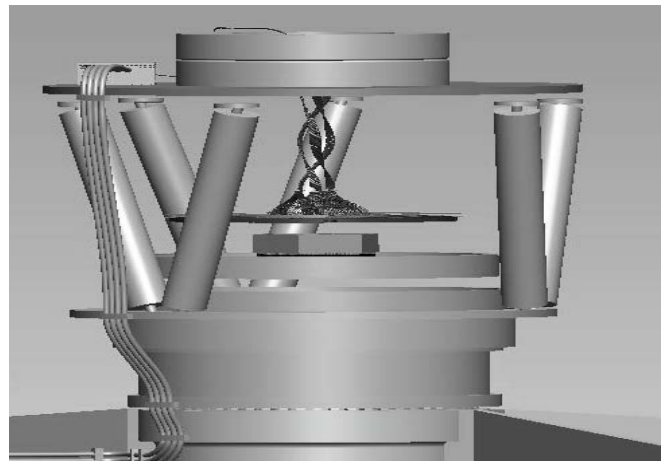

**Figure 13:** Concept for the plates system inside the chamber (PFS envelop) and the segmented tubes secured for the assembly plate by holders terminals.

**Figure 14:** PFI Chamber drawing with internal details of the segmented tubes incasing the optical fibers. The structure will be projected to have the cable wrap on the top.

The Cable C needs to be integrated with the COBRA system, which is a set of devices including COBRA motors that are responsible for the patrol of the fiber. The plan for the construction of this cable has 2 parts: the first one will be at LNA and, the second one, at JPL. The first part, Figure 15, includes: cutting the fibers, populate the arms with optical fibers, to glue the fibers, polish the fiber arms and to glue the microlens at the fiber extremity. The second part, Figure 16, includes: passing the optical fibers through the COBRA motors, populate the segmented tubes, passing the fibers through the connectors, and polishing of the connectors.

## Fabrication Process Cable C before send to JPL

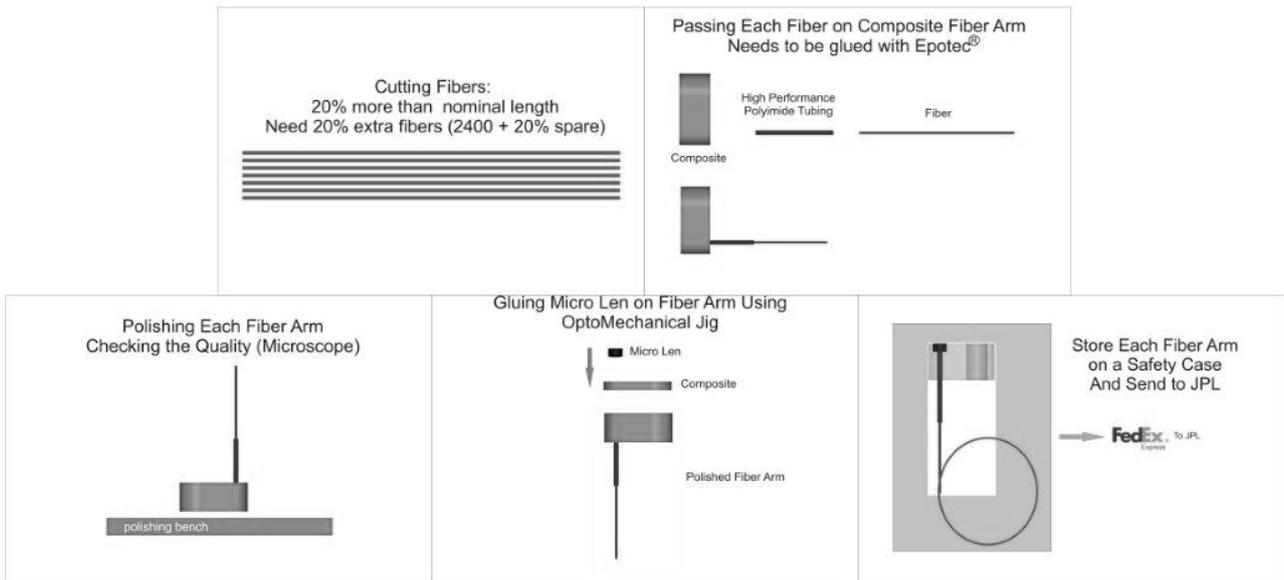

Figure 15: Fabrication process of the cable C before sends it to JPL

## Fabrication Process Cable C on JPL

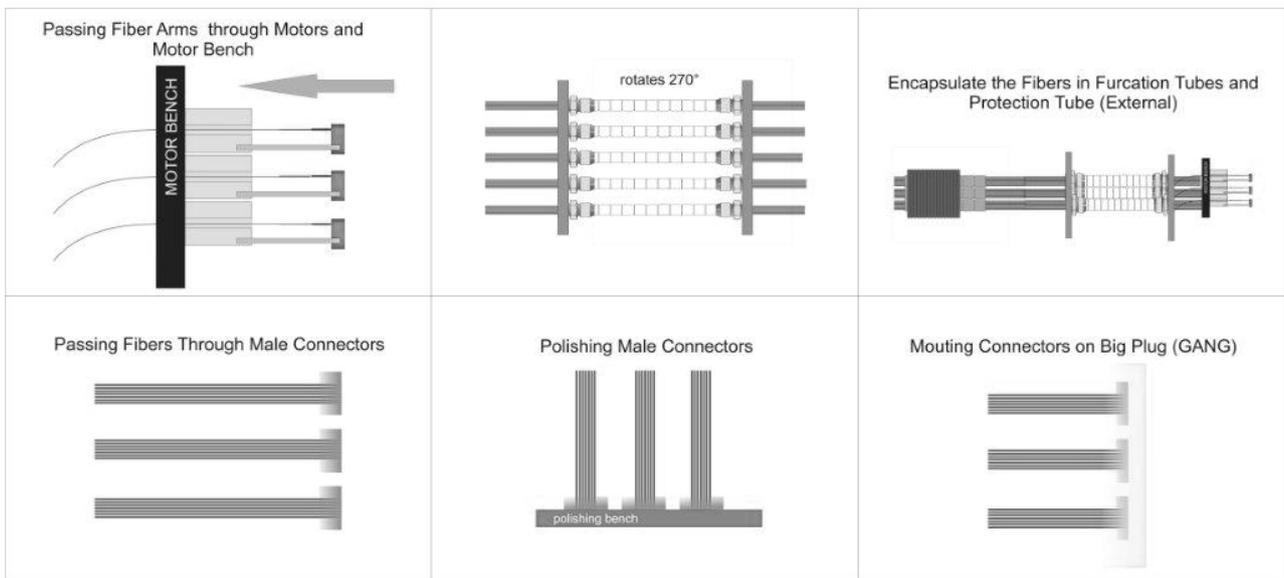

**Figure 16:** Fabrication process of the cable C at JPL

### 2.3.1 Composite and Fiber Arm

A fiber arm device was developed to be the extremity of the patrol device physically coupled at the COBRA motor device. The Figure 17 shows a sample of composite machined to form the "fiber arm" device for the "fiber positioner" system. The ideal condition requires a material with elasticity controlled so as not to cause stress or shift the positioning of the optical fiber under temperature gradients. Just for such purposes, we have developed a special composite formed from a mixture of EPO-TEK 301-2 and Zircon Oxide and others materials in nano-particle form, cured and submitted to a customized thermal treatment, Cesar et al. [01]. To avoid bubbles and points of stress, this mixture needs to be prepared in a separate receptacle inside a vacuum chamber. The resulting material is more resistant and harder than EPO-TEK 301-2 and it is found to be well suited to the fabrication of optical fiber arrays. An important secondary characteristic is the ease with which it can be polished. This feature is a result of the micro particles, which keep the polished surface very homogeneous during the final polishing procedure. The resulting composite combines the beneficial characteristics of both the epoxy and the zircon oxide. Its main property is the coefficient of thermal expansion, significantly lower than simple solidified epoxy; the exact value depends on the relative concentrations. While the characteristics of this particular composite still are under study, we have nevertheless deployed this material in the construction of devices for several fiber instruments. In order to slow down the F-ratio after the wide-field corrector by a factor of 1.28, microlens will be glued at the extremity of the fiber in the fiber arm, as shown in Figure 18. Accordingly, the F-ratio changes as follows: - Field center: F/2.18 F/2.79 - Field edge; F/2.01 F/2.57. Manufacturing viable alternatives are being investigated to produce this microlens, Figure 19 in a number of companies. At this moment, we are experimenting alternative techniques to align, assemble and glue the microlens at the fiber arm, Figure 20. The best option will be the most efficient and the faster.

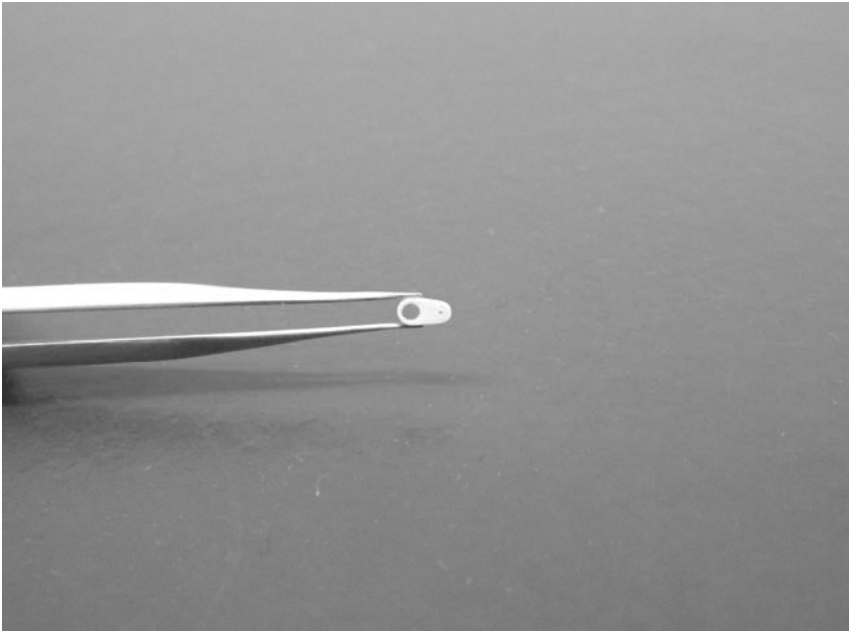

**Figure 17:** Fiber arm machined in composite. It shows a fiber arm device constructed with the composite to form the optical fiber ferrule of a fiber positioner system. This is a part of a complex patrol system that may be used as the positioner for WFMOS. The big advantage to the use of the composite is in the speed and quality of the polishing.

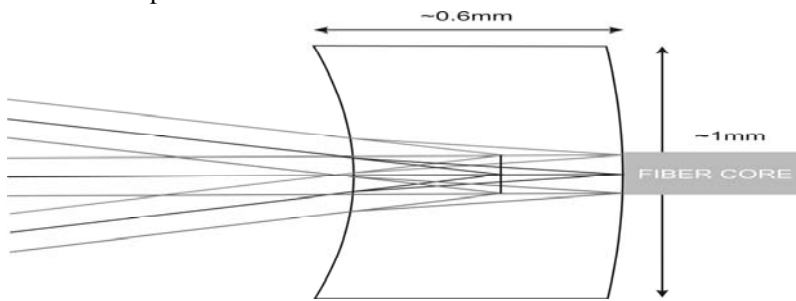

**Figure 19:** Lens design in evaluation

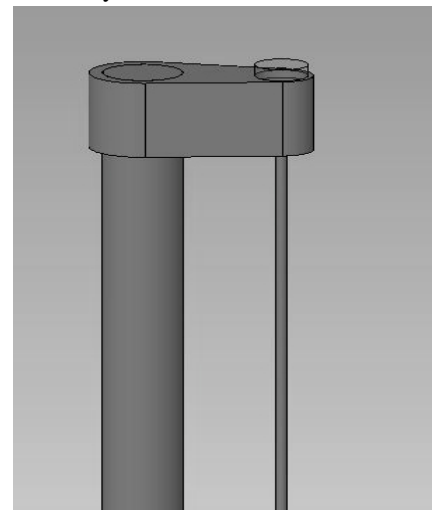

**Figure 18:** Fiber arm with optical fiber and lens

• How to align and assemble microlens and fiber:

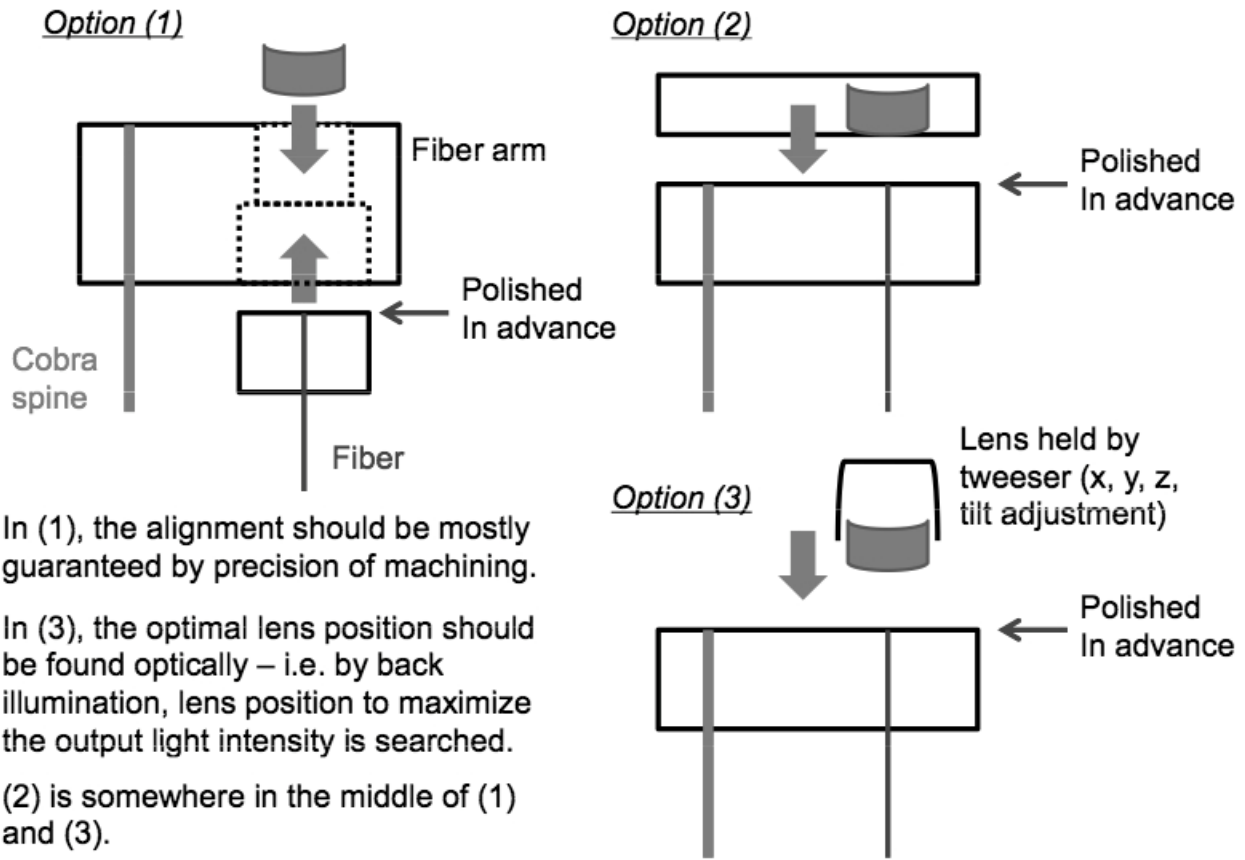

Figure 20: Three possible assemble process of the lens at the fiber arm

## 3. OPTICAL FIBERS IN TEST

The only fiber in test at this moment is the Polymicro FBP120170190. However, Fujikura Company is working with a possible different fiber that we may do tests in the near future. This will be a second option of optical fiber and all the tests described below will be repeated as soon as we have an available sample. We have prepared two samples with 6 and 50 meters length. Both extremities were encapsulated inside a composite ferrule and polished with high performance. We are avoiding measure small pieces like one or two meters because the effects of the annular degeneration.

### 3.1 Throughput & FRD Measurements

To measure the FRD properties of an optical fiber it is necessary to illuminate the fiber test with an input beam of known focal ratio. Then the output beam can be measured and compared with the input beam. We have adapted a method described by Barden, to measure absolute efficiencies.[02] In our experimental set-up; we can change the focal ratio of the input fiber in test. However, the most important was to feed the fibers around f/2.3 matched to the output focal ratio of Subaru's Hyper-Supreme corrector. The experimental apparatus used to achieve this is illustrated in Figure 21. A halogen lamp which light is diffused and condensed to illuminate a pinhole device is the primary light source. A filter centered at 550 nm defines the wavelength in which we are working on. The pinhole is the secondary light source and it projects light to be collimated using an achromatic collimator lens device. The collimated beam is then coupled into the test fiber by a scientific grabber CCD (Andor Solis DL-604 M-OEM). An iris diaphragm, placed in the collimated beam, can be used to select the input focal ratio. A CCD for alignment is fed by a beam splitter and allows the image of the source fiber to be accurately positioned with respect to the test fiber by means of observation on a TV monitor. A displacement device, defines the alignment of the optical axis of the CCD detector with either the output fiber with either the output of the collimated beam before at the test fiber entrance. In another words this tip-tilt translation stage allows obtain image from the output of the test fiber and image from the collimator focus. In fact, a defocus of 7 mm ± 0.01 mm was then applied to the CCD detector in both positions to avoid the saturation of the pixels. The Figure 22 shows the experiment assembled on the optical bench. The optical fibers can then be examined over a range of focal ratios. Background exposures were also taken for subtraction from the test exposures to remove the effects of hot pixels, stray light, etc. In general, the technique to measure FRD uses **a science-grade** CCD (with frame grabber) and employs the IRAF routines QPHOT and PPOFILE (Tody, 1993) for data reduction.[03]

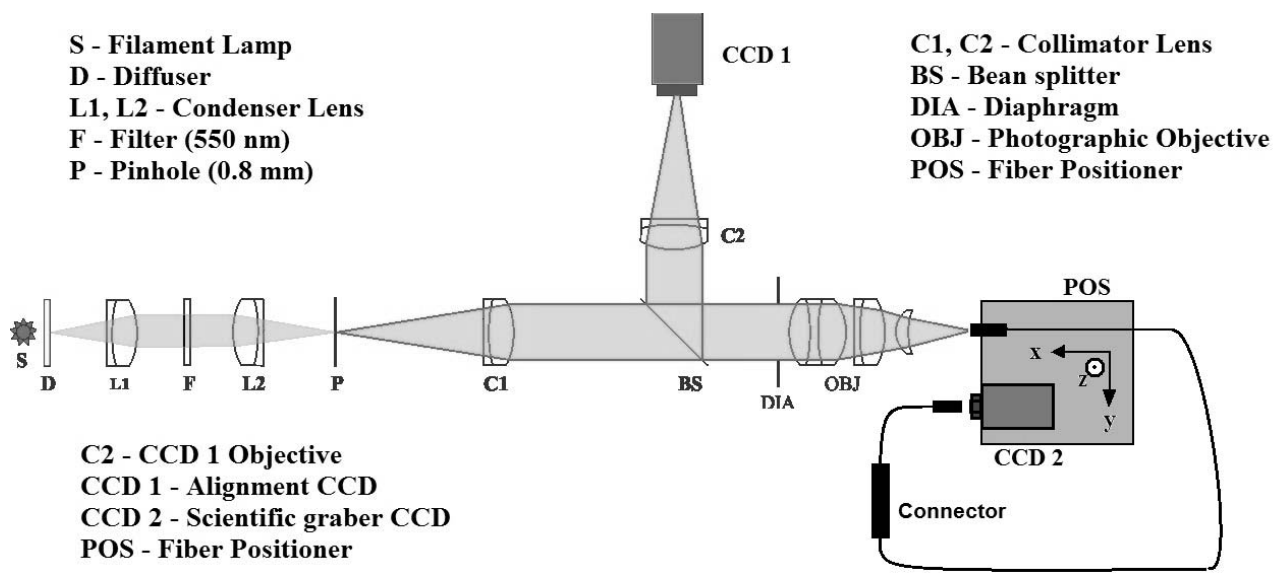

**S - Filament Lamp**
**D - Diffuser**
**L1, L2 - Condenser Lens**
**F - Filter (550 nm)**
**P - Pinhole (0.8 mm)**

**C1, C2 - Collimator Lens**
**BS - Bean splitter**
**DIA - Diaphragm**
**OBJ - Photographic Objective**
**POS - Fiber Positioner**

**C2 - CCD 1 Objective**
**CCD 1 - Alignment CCD**
**CCD 2 - Scientific graber CCD**
**POS - Fiber Positioner**

**Figure 21:** Schematic diagram used to measure absolute transmission

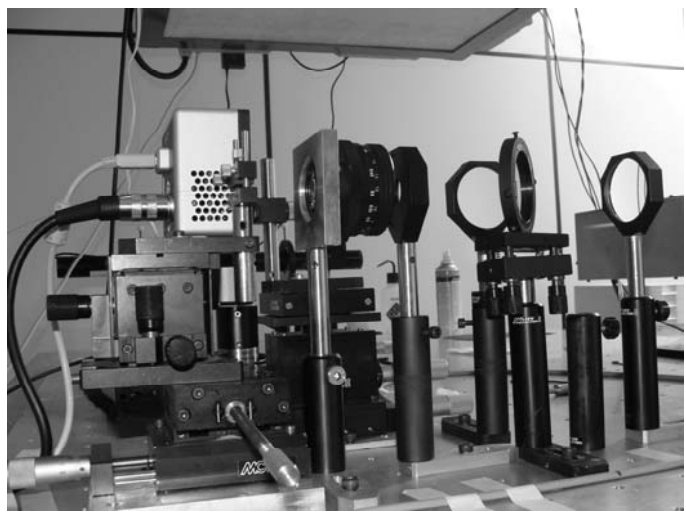

**Figure 22:** Optical bench to measure FRD Experimentation

Characteristics of the experimentation:

• Five Input F-ratios: F/2.2 – F/2.4 – F/2.6 – F/2.8 – F3.0
• Output graph curves between F/2 and F/4
• Two samples length: 6m and 50m
• Spotlight projected on the fiber from a "far field" source covering 90% of the core disc
• Software for data reduction Calculation *Flat Field and Dark Field
• Experimental error less than 1 %

## 3.2 Analysis Software

We have developed a custom software package (DEGFOC 3.0) to reduce the fiber images and to obtain throughput energy curves. This software works in a Windows[©] environment. We found it to be an effective solution for use in the optical laboratory environment allowing for ease of analysis. The DEGFOC 3.0 package gives curves of Enclosed Energy and Absolute Transmission as is shown in the Figure 23 with the option to save the result in ASCII format to be used in any graphic software. Fiber throughputs are automatically determined as a function of output focal ratio.

The first step of a measurement is an estimation of the background level to be subtracted from the test exposures to remove the effects of hot pixels and stray light. The software then finds the image center by calculating the weighted average of all pixels. It associates a radius with each pixel and calculates the eccentricity that, in the ideal case, should be zero. Our target here is to obtain the absolute transmission of the fiber at a particular Input Focal Ratio. After establishing the distance between the fiber test and the CCD detector, the software defines concentric annulus centered on the fiber image. These are then used to define the efficiency over a range of f-numbers at the exit of the fiber, where each f-number value contains the summation of partial energy emergent from the fiber. Each energy value is calculated by the number of counts within each annulus divided by total number of counts from the lens camera images with some small defocus. The limiting focal ratio that can propagate in the tested fiber is approximately F/2.2 taking in account the numerical aperture of this fiber to be $0.22 \pm 0.02$. Therefore we have defined F/2.2 to be the outer limit of the external annulus within which all of the light from the test fiber will be collected. The corresponding diameters of the annulus are converted to output focal ratios, multiplying them by the appropriate constant given by the distance between the fiber output end and the detector.

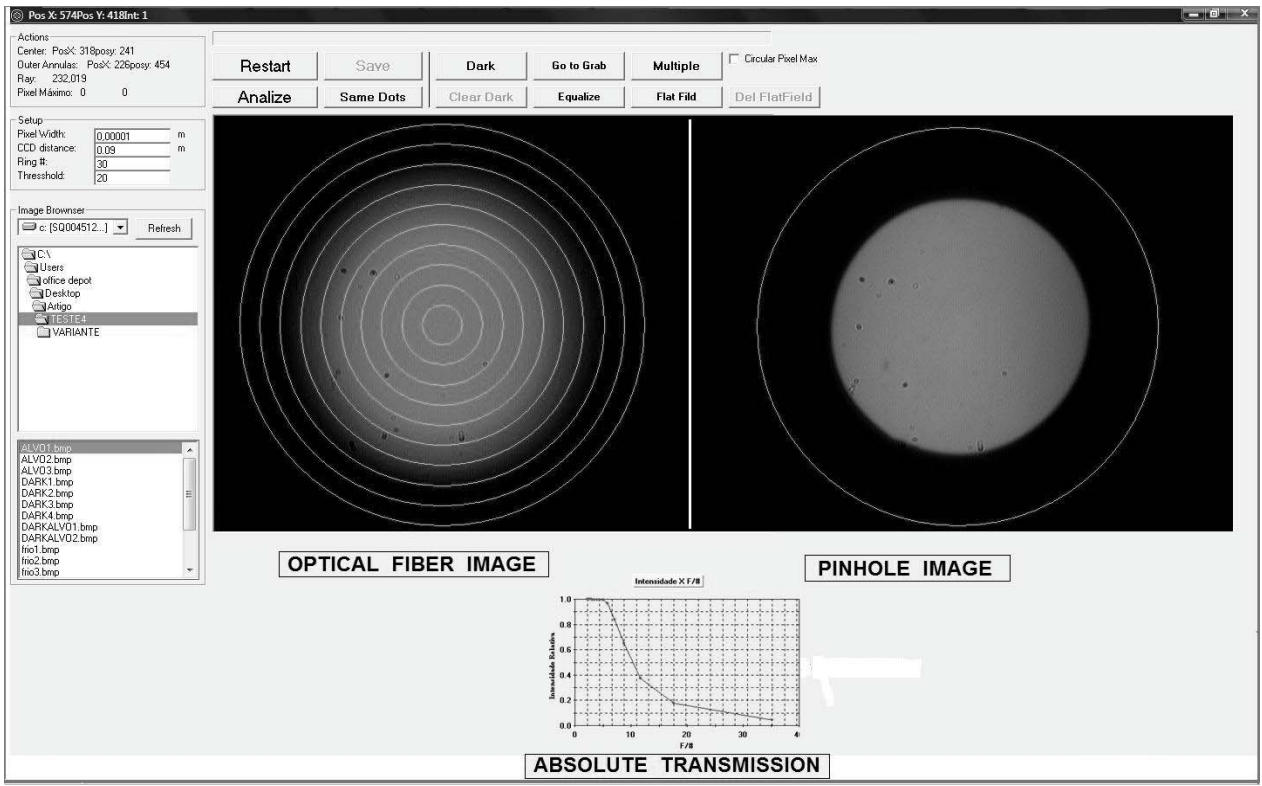

**Figure 23:** Print screen of the windows to the DEGFOC software. The software is user-friendly and has all the tools to produce the curve of enclosed energy as shown at the right. The center of the fiber spot, at left, is calculated from the weighted average of all pixels. Parameters such as distance between CCD and fiber, size of pixel and annulus number may be changed in the configuration box to optimize the calculation.

### 3.3 Results

Plots of absolute transmission versus output focal ratio for a sample with 6 meters Polymicro (FBP are presented in the Figure 24. For this sample we have obtained 6 graphs (Absolut Transmission & Output Focal Ratio), with 6 different Input Focal ratio; F/2.2, F/2.4, F/2.6, F/2.8 and F/3.0. The analysis has shown us that there is a box with 3 regions to be explored in operational conditions. A gray code, Bad, Good and Nice, despite to be arbitrary, may facilitate the analysis of the general throughput of the instrument. For example, if we intend to feed the fibers in Cable C around 6 meters length with F/2.6 would be reasonable expect obtain efficiency around 80%.

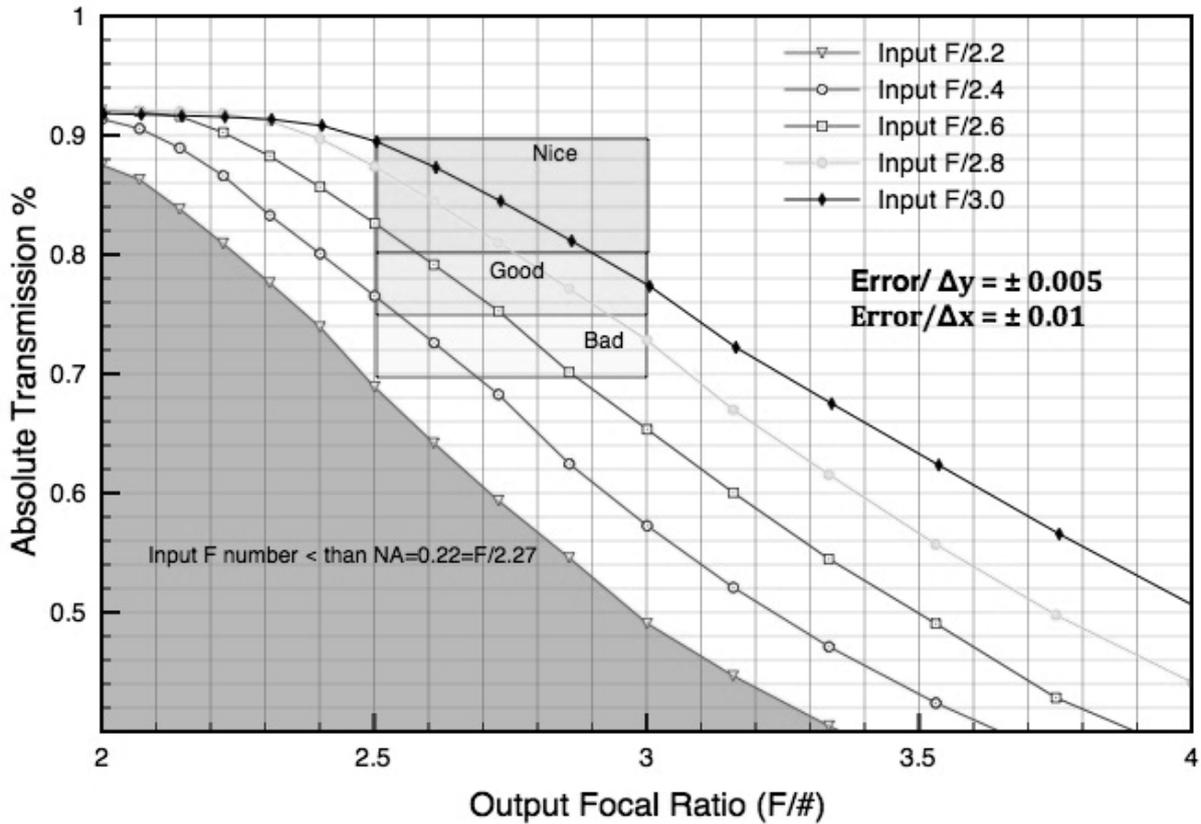

**Figure 24:** Optical Fibers Investigation FRD in Absolute Transmission Curves - 6 meters

## 4. MULTI-FIBERS CONNECTORS SYSTEM

The multi-fibers connector under study, shown in Figure 25, is produced by USCONEC Company. The called Apogee connector can connect 32 optical fibers in a single ferrule. The material of this ferrules are manufactures durable, composite, Polyphenylene Sulfide (PPS) based thermoplastic ferrules. The connections are held in place by a push-on/pull-off latch, and the connector can also be distinguished by a pair of metal guide pins that protrude from the front of the connector. Two fibers per connector will be used to monitoring the connection procedure. Lifetime and throughput are being investigated at this moment, but was found to be easy to polish and it is small enough to be mounted in groups. Moreover, its use in the spectrograph Apogee (SDSS) has produced excellent results.[04]

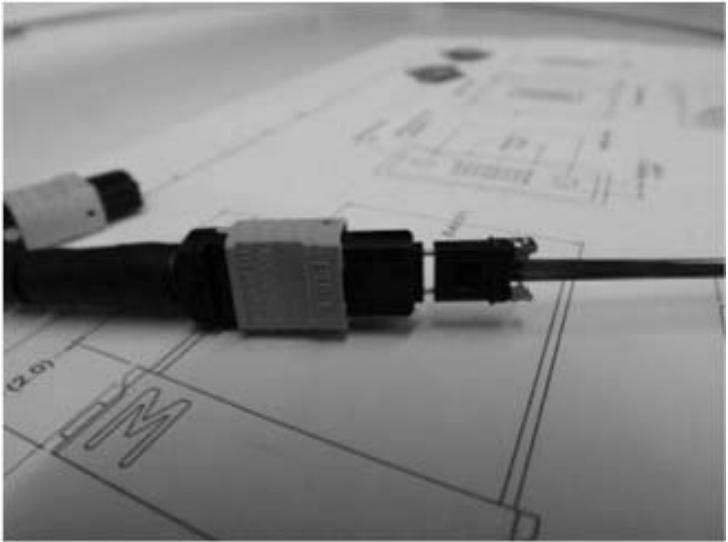

It will be necessary to use two different types of structures to organize the connectors: Tower Connector system, with 80 connectors will be a group of connector's cells, TBD, to connect cable B (Telescope Structure) with cable C (Positioners Plate), Figure 26. Gang Connector system, is a group of 8 gang connectors, each one with 10 Apogee connectors to connect cable B (Telescope Structure) with cable A (Spectrograph), Figure 27. The bench tests with the Apogee connector and the chosen fibers should measure the throughput of light and the stability after many connections and disconnections. The test proceeding to evaluate the throughput is in developing at this moment. The lifetime of the ferrules should also be studied in the next few months.

**Figure 25:** 32 optical fibers multi-connector produced by USCONEC Company in test phase.

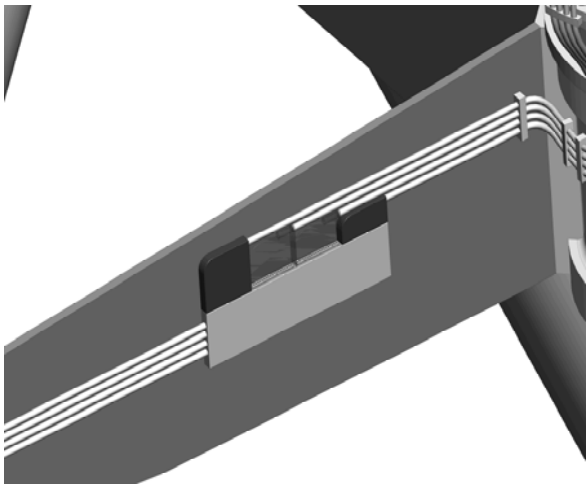

**Figure 26:** Tower Connector System, to be used at the telescope top end side. The Tower Connector system will be a group with 80 USCONEC multi-connectors

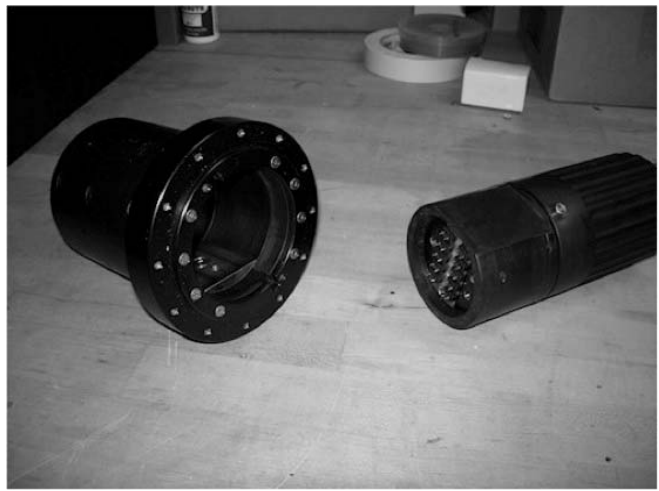

**Figure 27:** Gang Connector System, to be used at spectrograph side. Ten multifibers connectors inside 30 x 10 = 300 ➔ 2 Gang Connector per pseudo-slit

## 5. SUMMARY AND CONCLUSIONS

We described here the conceptual design for FOCCoS, (Fiber Optical Cable and Connectors System) to be a subsystem of SuMIRe- PFS, (SUBARU Measurement of Images and Redshifts- Prime Focus Spectrograph). A set of 4 pseudo-slits will comprise one of the extremities of the cable system. The other extremity will direct the fibers for fiber arms devices as the part of a patrol system. Each fiber arm will have one end of optical fiber polished and coupled at a microlens. FOCCoS is being projected for work with 2400 optical fibers segmented in 3 cables connected by sets of multi-fibers connectors. The multi-fibers connector under study is produced by USCONEC Company and Polymicro Company produces the optical fiber actually in test. It is expected to experiment with other types of optical fiber produced by Fujikura Company. A new composite material with optimized characteristics will be used as part of the substrate of the pseudo-slits and the fiber arms device.

# 6. ACKNOWLEDGMENTS

We gratefully acknowledge support from: The Funding Program for World-Leading Innovative R&D on Science and Technology. SUBARU Measurements of Images and Redshifts (SuMIRE), CSTP, Japan. Fundação de Amparo a Pesquisa do Estado de São Paulo (FAPESP), Brasil. Laboratório Nacional de Astrofísica, (LNA) e Ministério da Ciência Tecnologia e Inovação, (MCTI), Brasil. We would like also to gratefully acknowledge Youichi Ohyama and Hung-Hsu Ling, PO members.